\documentclass[twocolumn,amssymb,amsmath,aps,floatfix,prl,superscriptaddress]{revtex4-1}

\usepackage{hyperref}
\usepackage[all]{hypcap}
\usepackage{array}
\usepackage{graphicx}
\usepackage{amsmath,amsthm,amssymb}

\usepackage{color}

\begin{document}

\title{$Z^0$+jet correlation with NLO-matched parton-shower and jet-medium interaction in high-energy nuclear collisions}

\author{Shan-Liang Zhang}
\affiliation{Key Laboratory of Quark and Lepton Physics (MOE) and Institute
of Particle Physics, Central China Normal University, Wuhan 430079, China}

\author{Tan Luo}
\affiliation{Key Laboratory of Quark and Lepton Physics (MOE) and Institute
of Particle Physics, Central China Normal University, Wuhan 430079, China}

\author{Xin-Nian Wang}
\affiliation{Key Laboratory of Quark and Lepton Physics (MOE) and Institute
of Particle Physics, Central China Normal University, Wuhan 430079, China}
\affiliation{Nuclear Science Division Mailstop 70R0319,  Lawrence Berkeley National Laboratory, Berkeley, CA 94740}

\author{Ben-Wei Zhang\footnote{bwzhang@mail.ccnu.edu.cn}}
\email{Corresponding author: bwzhang@mail.ccnu.edu.cn}
\affiliation{Key Laboratory of Quark and Lepton Physics (MOE) and Institute
of Particle Physics, Central China Normal University, Wuhan 430079, China}

\begin{abstract}
The impact of jet quenching on $Z^0$-tagged jets in relativistic heavy-ion collisions at the Large Hadron Collider (LHC) is investigated.  We employ Sharpa Monte Carlo program that combines next-to-leading order matrix elements with matched resummation of parton shower to compute the initial $Z^0$+jet production. The Linear Boltzmann Transport (LBT) model is then used to simulate the propagation, energy attenuation of  and medium response induced by jet partons in the quark-gluon plasma.
With both higher-order corrections and matched soft/collinear radiation as well as a sophisticated treatment of parton energy loss and medium response in LBT,
our numerical calculations can provide the best description so far of all available observables of $Z^0$+jet simultaneously in both p+p and Pb+Pb collisions, in particular, the shift of the distribution in transverse momentum asymmetry $x_{\rm jZ}=p_T^{\rm jet}/p_T^Z$, the modification of azimuthal angle correlation in $\Delta\phi_{\rm{j}Z}=|\phi_{\rm jet}-\phi_Z|$ and the overall suppression of average number of $Z^0$-tagged jets per boson $R_{\rm jZ}$ at $\sqrt s =5.02$ TeV as measured by the CMS experiment. We also show that higher-order corrections to $Z^0$+jet production play an indispensable role in understanding $Z^0$+jet azimuthal angle correlation at small and intermediate $\Delta\phi_{\rm jZ}$, and momentum imbalance at small $x_{\rm jZ}$. Jet quenching of the sub-leading jets is shown to lead to suppression of $Z^0$+jet correlation at small azimuthal angle difference $\Delta\phi_{\rm jZ}$ and at small $x_{\rm jZ}$.
\end{abstract}
\pacs{25.75.Bh,25.75.Ld, 24.10.Lx}
\maketitle

{\bf \textit{Introduction}} ---
Jet quenching or suppression of energetic partons due to energy loss in medium has long been proposed to probe properties of the quark-gluon plasma (QGP) in heavy-ion collisions (HIC)~\cite{Wang:1991xy,Gyulassy:2003mc, Qin:2015srf, Vitev:2008rz, Vitev:2009rd, Qin:2010mn, CasalderreySolana:2010eh,Young:2011qx,He:2011pd,ColemanSmith:2012vr,Zapp:2012ak,Ma:2013pha, Senzel:2013dta, Casalderrey-Solana:2014bpa,Milhano:2015mng,Chang:2016gjp,Majumder:2014gda, Chen:2016cof, Chen:2016vem, Chien:2016led, Apolinario:2017qay,Connors:2017ptx}. Gauge-boson-tagged jet production is regarded as a ``golden channel" to study the jet quenching
~\cite{ Wang:1996yh,Qin:2009bk}. The boson will not participate in the strong-interactions directly and can be considered as the proxy of the initial energy of the parton before it propagates through the QGP medium and loses energy~\cite{Dai:2012am,Wang:2013cia,Chen:2018fqu}.
Though jet production associated with a direct photon in HIC has already been accessible at the Relativistic Heavy-ion Collider (RHIC), the unprecedented energies available at the Large Hadron Collider (LHC) open a new window for $Z^0$-tagged jet production in HIC,  where the $Z^0$ gauge boson not only escapes the QGP medium unattenuated, but is also free from fragmentation processes due to its very large mass.


Recently CMS Collaboration has reported the first measurement of $Z^0$-tagged jet production  in both p+p and Pb+Pb collisions  at $\sqrt s=5.02$ TeV at the LHC~\cite{Sirunyan:2017jic}.   Though the CMS data on $Z^0$+jet in Pb+Pb collisions can be qualitatively described by several  theoretical models, such as GLV~\cite{Neufeld:2010fj, Neufeld:2012df, Kang:2017xnc},  Hybrid model~\cite{Casalderrey-Solana:2015vaa} and JEWEL~\cite{KunnawalkamElayavalli:2016ttl},
it is still a challenge to quantitatively describe all the available experimental observables of $Z^0$+jet simultaneously and their p+p baseline by simulations based on a leading order (LO) matrix element (ME) matched parton shower (PS) event generator. The $Z^0$+jet azimuthal angle correlation $\Delta \phi_{\rm jZ}= |\phi_{\rm jet}-\phi_{Z}|$ and the distributions in average number of $Z^0$-tagged jets  $R_{\rm jZ}=N_{ \rm jZ}/N_{Z}$  are in particular very sensitive to QCD higher-order corrections~\cite{Chatrchyan:2013tna, Sirunyan:2017jic}. It is therefore of a great advantage to use the next-to-leading order (NLO) pQCD computations of hard scattering complemented with resummation of soft/collinear parton shower and the state of the art simulations of parton propagation in QGP medium in the study of $Z^0$-jet correlation in high-energy HIC.

In this Letter, we report the first numerical study with such a theoretical model:  the Monte Carlo program Sherpa \cite{Gleisberg:2008ta}, which combines the NLO pQCD with resummation of a matched PS, is used for the initial $Z^0$-tagged jet production and  provides an excellent description of $Z^0$+jet production in elementary p+p collisions; the parton propagation in QGP medium is simulated by the Linear Boltzmann Transport (LBT) model \cite{Wang:2013cia,He:2015pra,Cao:2016gvr} with bulk medium evolution provided by the Berkeley-Wuhan CLVisc 3+1D hydrodynamics \cite{Pang:2012he,Pang:2014ipa}. We refer this model as NLO+PS LBT model. We will confront our results with available data for all four observables of $Z^0$+jet in both p+p and Pb+Pb collisions: azimuthal correlation $\Delta\phi_{\rm jZ}$, $p_T$ asymmetry $x_{\rm jZ}$ distribution and its mean value $ \langle x_{\rm jZ}\rangle$,  as well as the average number of associated jets per $Z^0$ boson $R_{\rm jZ}$. We will focus in particular on effects of multiple jets associated with $Z^0$ and their suppression on the azimuthal correlation and $p_T$ asymmetry in Pb+Pb collisions.


{\bf \textit{Sherpa}} --- Sherpa is a complete Monte Carlo event generator that  simulates all high-energy reactions between particles in the Standard Model.  Sherpa employs several emerging approaches~\cite{Hoeche:2009rj,Hoche:2010kg,Hoeche:2012yf} which provide NLO ME matched  to the resummation of the Collins-Soper-Sterman~\cite{Collins:1984kg} dipole PS~\cite{Schumann:2007mg,Gleisberg:2007md} to calculate low jet multiplicities and LO matched parton shower to simulate high jet multiplicities.
The matching scheme can be formulated symbolically as:
\begin{eqnarray}
&\langle O\rangle^{(NLO+PS)} =  \int d\Phi_B \left[ B+\widetilde{V}+I^{S} \right](\Phi_B)\widetilde{PS}_B(\mu_Q^2,O)  \nonumber \\
&+ \int d\Phi_R \left[ R(\Phi_R)-D^{S}(\Phi_B*\Phi_1)\right]\widetilde{PS}_R(t_R,O) .
\label{nloPS}
\end{eqnarray}
 where $ \Phi_B$ is the Born phase-space, $ \Phi_R$ is real phase-space, B, $\widetilde{V}$ and R denote Born, virtual and real ME respectively.  $D^S$ is the subtraction term which has the same soft divergence as the real terms in the subtraction scheme; $I^{S}=\int d\Phi_1   D^{S}$  is the integrated subtraction term. The introduction of $D^S$ and $I^{S}$ makes both matrix element part finite. $\widetilde{PS}_B(\mu_Q^2,O) $ and $\widetilde{PS}_R(t_R,O)$ is the parton shower branch for Born phase and real corrected phase space respectively, with  $\mu_Q^2$  and $t_R$ the shower starting points~\cite{Collins:2000qd,Chen:2001ci,Hoeche:2011fd}. In our simulations, the OpenLoops program\cite{Cascioli:2011va} calculates  loop-level diagram elements while Sherpa calculates tree-level diagram elements\cite{Krauss:2001iv,Gleisberg:2008fv} and makes phase space integration with the parton density set ``CTEQ14nlo".

\begin{figure}[tbp]
\begin{center}
\hspace*{-0.2in}
\vspace*{-0.1in}
\includegraphics[width=4.5in,height=4.0in,angle=0]{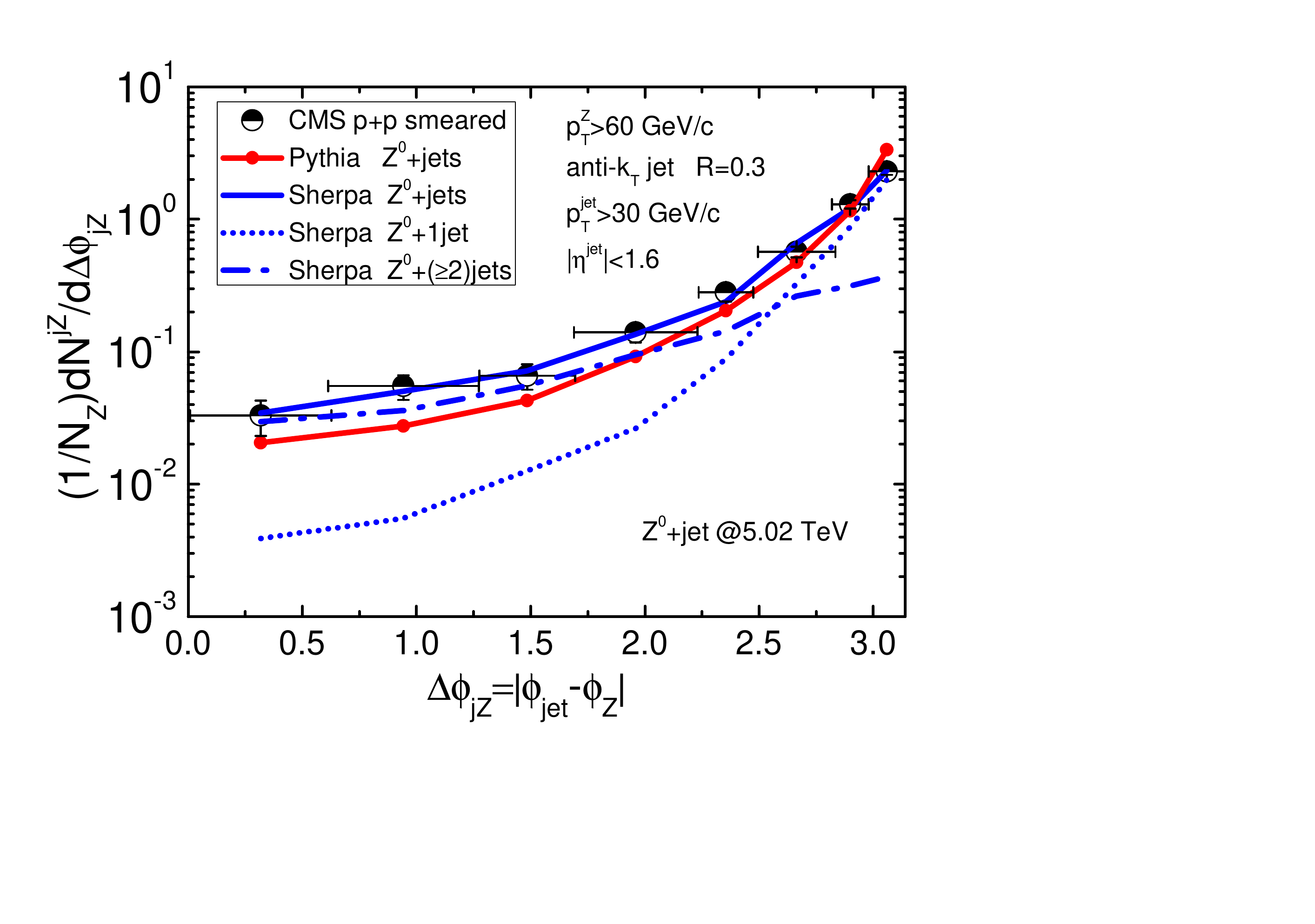}
\end{center}
\vspace{-10pt}
\caption[*]{(Color online) Comparison between the azimuthal angle correlation $\Delta \phi_{\rm jZ}$ of $Z^0$+jet by CMS data~\cite{Sirunyan:2017jic} and theoretical simulations of Sharpa (Blue) and Pythia (Red) in p+p collisions at $\sqrt s = 5.02$ TeV. The doted (the dash-dotted) line shows the contribution from $Z^0+1$jet ($Z^0+(\ge 2)$jets).}\label{zjetdeltaphi}
\end{figure}

We show in Fig.~\ref{zjetdeltaphi} the $Z^0$+jet correlation in azimuthal angle $\Delta \phi_{\rm jZ}$  in p+p collisions simulated by Sherpa as compared to the default Pythia 6.4 result and CMS data~\cite{Sirunyan:2017jic}. The Sherpa p+p baseline result shows an excellent agreement with experimental data, while Pythia 6.4  slightly overshoots the azimuthal distribution at large  $\Delta \phi_{\rm jZ} \sim \pi$  and significantly underestimates the distribution  by a factor of $\sim 2$ at small  $\Delta \phi_{\rm jZ}$.  Contributions from $Z^0$+1 jet and $Z^0$+($\ge 2$) jets to the azimuthal correlation in p+p collisions from Sherpa are also shown in Fig.~\ref{zjetdeltaphi}.  Contributions from $Z^0$+($\ge 2$) jets from NLO processes are much broader than that of $Z^0$+1 jet and dominate the distribution at small $\Delta \phi_{\rm jZ}$ region.  $Z^0$+1 jet processes contribute mostly in large $\Delta \phi_{\rm jZ}$ region where soft/collinear radiation from PS dominates.

To obtain the above numerical results and in the rest of this Letter, we adopt the kinematic cuts by CMS experiment~\cite{Sirunyan:2017jic} to select $Z^0$-tagged jets in both p+p and Pb+Pb collisions. For $Z^0\rightarrow e^+e^-$ decay, electrons are required to have $p_{T}^{e}>20 $ GeV, $|\eta^{e}|<2.5$ and are excluded in the kinematic region $1.44<|\eta^{e}|<2.47$.  For $Z^0 \rightarrow \mu^+\mu^-$ decay, kinematic cuts for muons are $p_{T}^{\mu}>10 $ GeV, $|\eta^{\mu}|<2.4$. $Z^0$ bosons are reconstructed by opposite-charge electron or muon pairs, with reconstructed mass  $70<M_{ll}<110$ GeV, and transverse momentum $p_{T}^{Z}>40$ GeV. Jets are constructed by FASTJET~\cite{Cacciari:2011ma}  from final partons with the anti-$k_T$ algorithm~\cite{Cacciari:2008gp} and jet cone size $R\equiv\sqrt{(\Delta\phi)^2+(\Delta y)^2}=0.3$. We have neglected the effect of hadronization. All the jets tagged by a boson should pass thresholds of $ p_{T}^{\rm jet}>30$ GeV, $|\eta^{\rm jet}|<1.6$,  and are rejected in a cone of $R <0.4 $ from a lepton to reduce jet energy contamination.

{\bf \textit{LBT Model}} ---In this study, propagation of fast partons in hot QGP is simulated within the LBT model \cite{Wang:2013cia,He:2015pra,Cao:2016gvr}  that includes both elastic and inelastic processes of parton scattering for both jet shower and thermal recoil partons in the QGP.  The elastic scattering is described by the linear Boltzmann equation~\cite{Wang:2013cia,He:2015pra,Cao:2016gvr},
\begin{eqnarray}
p_1\cdot&\partial f_a(p_1)=-\int\frac{d^3p_2}{(2\pi)^32E_2}\int\frac{d^3p_3}{(2\pi)^32E_3}\int\frac{d^3p_4}{(2\pi)^32E_4} \nonumber \\
&\frac{1}{2}\sum _{b(c,d)}[f_a(p_1)f_b(p_2)-f_c(p_3)f_d(p_4)]|M_{ab\rightarrow cd}|^2 \nonumber \\
&\times S_2(s,t,u)(2\pi)^4\delta^4(p_1+p_2-p_3-p_4),
 \end{eqnarray}
where $f_{i=a,b,c,d}$ are parton phase-space distributions, $|M_{ab\rightarrow cd}|$ is the corresponding elastic  ME.
$S_2(s,t,u)$ stands for a Lorentz-invariant regulation condition~\cite{Wang:2013cia,He:2015pra,Cao:2016gvr}.
The inelastic scattering is described by the higher twist formalism for induced gluon 
radiation~\cite{Guo:2000nz, Zhang:2003yn, Zhang:2003wk,Majumder:2009ge} as,
\begin{equation}
\frac{dN_g}{dxdk_\perp^2 dt}=\frac{2\alpha_sC_AP(x)\hat{q}}{\pi k_\perp^4}\left(\frac{k_\perp^2}{k_\perp^2+x^2M^2}\right)^2\sin^2\left(\frac{t-t_i}{2\tau_f}\right) \, ,
 \end{equation}
where $x$ and $k_\perp$ denote the energy fraction and transverse momentum of the radiated gluon,
 $ P(x)$ the splitting function, $\hat{q}$  jet transport coefficient, and
 $\tau_f=2Ex(1-x)/(k_\perp^2+x^2M^2) $ the formation time.  The information on local temperature and fluid velocity of the dynamically evolving bulk medium is provided by
 3+1D CLVisc hydrodynamical model \cite{Pang:2012he,Pang:2014ipa} with initial conditions from the AMPT model \cite{Lin:2004en} averaged over 200 events for each centrality. Parameters in the CLVisc are chosen to reproduce experimental data on bulk hadron spectra.  The only parameter in LBT model that controls the strength of parton interaction is strong coupling $\alpha_s$ which is chosen as $\alpha_s=0.20$ in this study for the best fit of the experimental data. LBT model has been used to describe successfully several important jet quenching observables, such as photon tagged hadron/jet production, light and heavy flavor meson suppression~\cite{Wang:2013cia,He:2015pra,Cao:2016gvr,Chen:2017zte,Luo:2018pto}.

\begin{figure}[tbp]
\begin{center}
\includegraphics[scale=0.4]{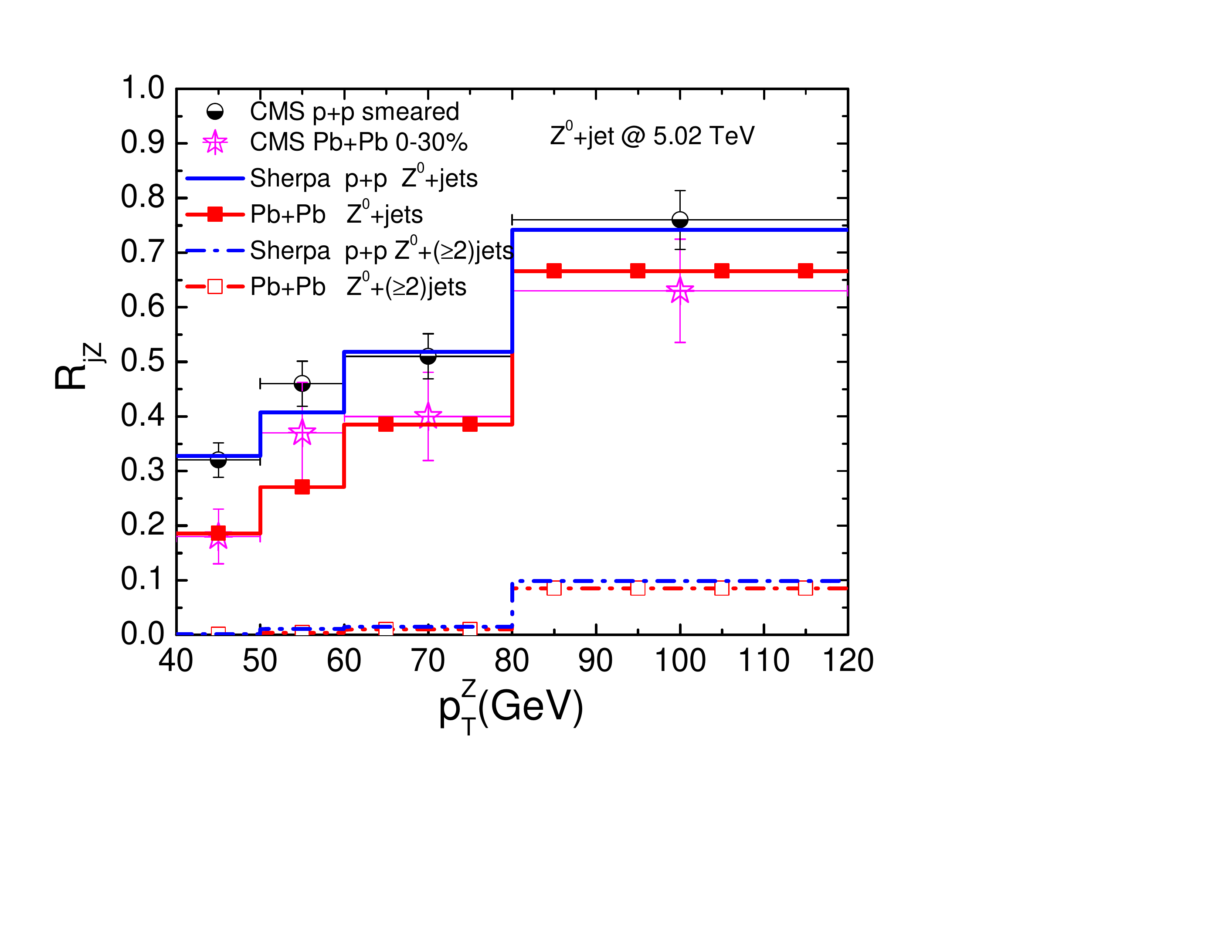}
\end{center}
\vspace{-20pt}
\caption[*]{(Color online) The calculated $R_{\rm jZ}$ distributions of $Z^0$+jet as a function of $p_T^Z$  in p+p (blue)  and Pb+Pb collisions (red) at $\sqrt s = 5.02$ TeV as compared to CMS data\cite{Sirunyan:2017jic}. The dash-dotted lines show contributions 
from $Z^0+(\ge 2)$jets. }
\label{zjet_Rjz}
\end{figure}


\begin{figure}[tbp]
\begin{center}
\includegraphics[scale=0.4]{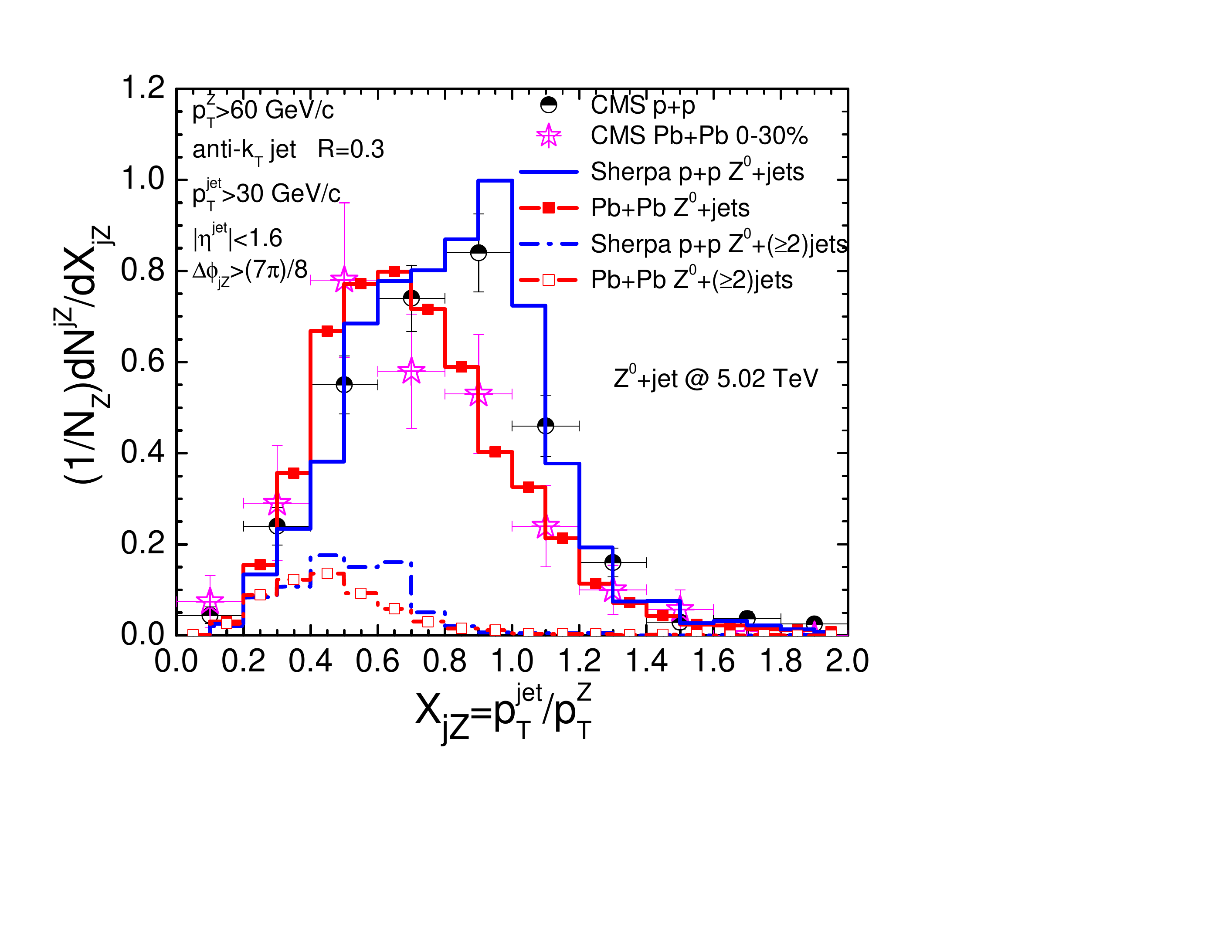}
\end{center}
\vspace{-20pt}
\caption[*]{(Color online) The calculated momentum imbalance of $Z^0$+jet in p+p (blue) and Pb+Pb collisions (red) at $\sqrt s = 5.02$ TeV  as compared to CMS data\cite{Sirunyan:2017jic}. The dash-dotted lines show the contributions from $Z^0+(\ge 2)$jets.}
\label{zjet_xjz}
\end{figure}

{\bf \textit{Results and Discussions}} --- Using Sherpa NLO+PS event generator and LBT model, we can study medium modification of $Z^0$+jet correlation in Pb+Pb at the LHC. Effects of cold nuclear matter are found to be rather small in the kinematics we are interested in~\cite{Ru:2014yma}. All partons, jet shower, radiated and medium recoil partons, are used for jet reconstruction with FASTJET. In the following calculations of $Z^0$+jet correlation in Pb+Pb, the underlying events background subtraction has been carried out following the procedure adopted by CMS experiment~\cite{UE-CMS}. No subtraction is applied in p+p results. The energy and azimuthal angle resolution of the detector are simulated by a Gaussian smearing with centrality-dependent parametrization as given by CMS experiment~\cite{Sirunyan:2017jic}.
\begin{figure}[tbp]
\begin{center}
\includegraphics[scale=0.4]{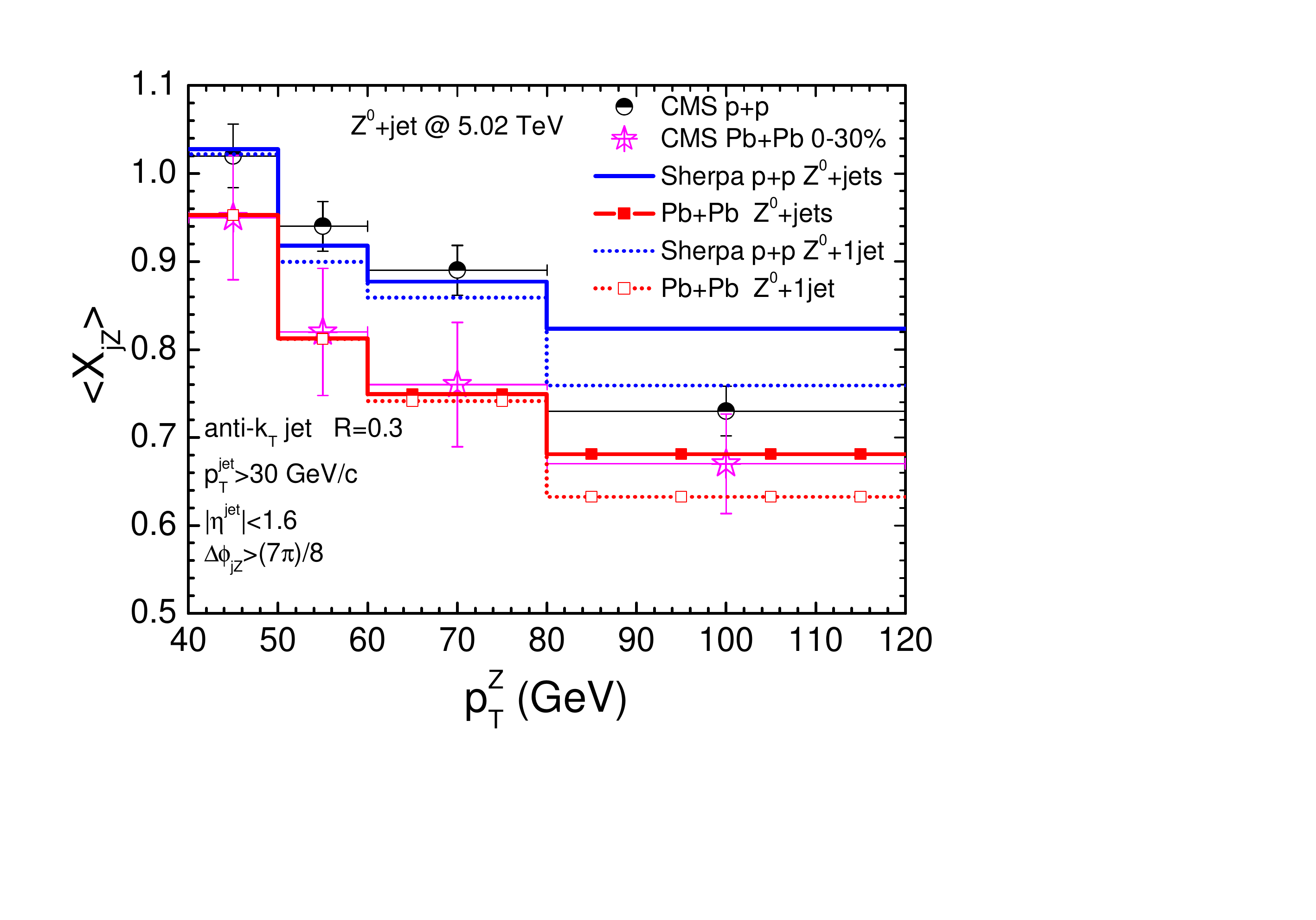}
\end{center}
\vspace{-20pt}
\caption[*]{(Color online) Numerical calculations on the mean value of $x_{\rm jZ}$  as a function of $p_T^Z$ in p+p (blue) and Pb+Pb collisions (red) collisions at $\sqrt s = 5.02$ TeV as compared to the CMS data~\cite{Sirunyan:2017jic}. The dotted lines show the contributions from $Z^0+1$jet. }\label{zjet_xjz_mean}
\end{figure}

The distribution in average number of tagged jets per $Z^0$ boson $R_{\rm jZ}= N_{\rm jZ}/N_{Z}$ is shown in Fig.~\ref{zjet_Rjz}.  We note that the jet selection threshold $ p_{T}^{\rm jet}>30$~GeV imposes a strong constraint on the phase space of $Z^0$-tagged jets.  Significant suppression for $R_{\rm jZ}$ is observed in Pb+Pb collisions relative to that in p+p collisions. This is a direct consequence of jet energy loss that shifts the final transverse momentum of a larger fraction of $Z^0$-tagged jets below the $p_T^{\rm Z}=30 $ GeV threshold.
The difference between $ R_{\rm jZ}$ in p+p and Pb+Pb changes slowly with $p_T^{\rm jet}$. We note that jets with high recoil $p_T$ associated with a $Z^0$ boson are dominated by quark jets. The contribution of $Z^0$+ multi-jets to $R_{\rm jZ}$ distribution is small in both p+p and Pb+Pb collisions because of the kinematical constraints of finding  multiple high-energy jets whose energy should not exceed half of that of $Z^0$ boson.


Fig.~\ref{zjet_xjz} shows our model calculations of the distribution in the transverse momentum asymmetry $x_{\rm jZ}=p_{T}^{\rm jet}/p_{T}^{Z}$ for $Z^0$-tagged jet at $\sqrt s = 5.02$ TeV in p+p and Pb+Pb collisions as compared with CMS data. A cut $\Delta \phi_{\rm jZ} > 7\pi/8 $ has been imposed to select the most back-to-back $Z^0$+jet pairs.  Compared to p+p collisions, the asymmetry distribution in $x_{\rm jZ}$ is broadened and shifted toward a smaller value of $x_{\rm jZ}$ in $0-30\%$ central Pb+Pb collisions due to jet energy loss in the QGP medium while the transverse momentum of $Z^0$ boson remains the same.  The distribution is dominated by $Z^0$+1 jet process at large $x_{\rm jZ}$, but has almost 50$\%$ contributions from higher order corrections at small $x_{\rm jZ} < 0.5$. For completeness we also show our model results on the mean value of momentum imbalance $\langle x_{\rm jZ}\rangle$ at different $p_T^Z$ bins in Fig.~\ref{zjet_xjz_mean}


%
%
%

\begin{figure}[tbp]
\begin{center}
\includegraphics[width=4.5in,height=4.0in,angle=0]{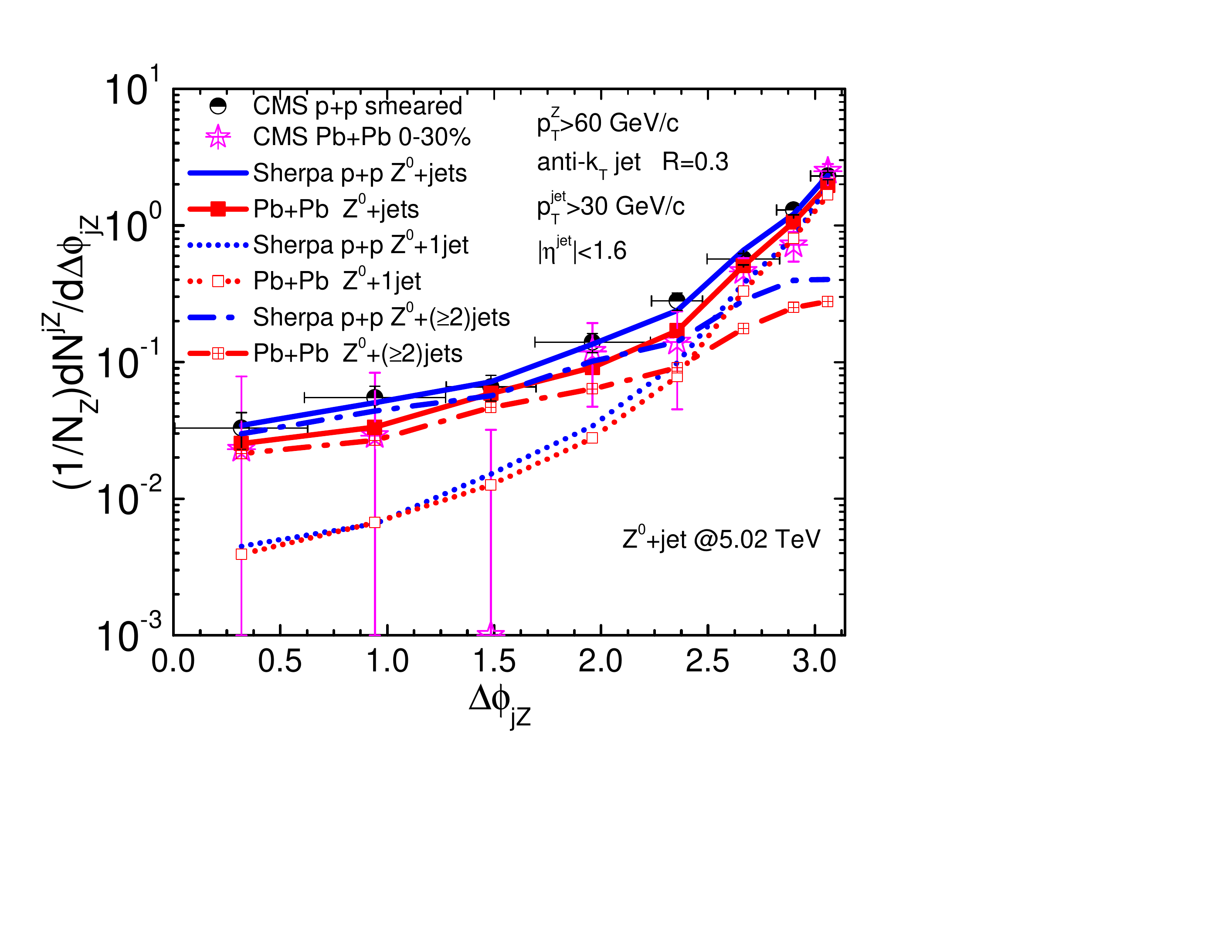}
\end{center}
\vspace{-20pt}
\caption[*]{(Color Online) Numerical results of the azimuthal angle correlation  in $\Delta \phi_{\rm jZ}$ in p+p (blue) and Pb+Pb collisions (red) at $\sqrt s = 5.02$ TeV  as compared to CMS data\cite{Sirunyan:2017jic}. The dotted (dash-dotted) lines show the 
contributions from $Z^0+1$jet ($Z^0+(\ge 2)$jets).}
\label{zjet_deltaphi}
\end{figure}

We show in Fig.~\ref{zjet_deltaphi} our calculations of $Z^0$+jet correlation in the azimuthal angle difference  $\Delta \phi_{\rm jZ}$ between  $Z^0$ boson and jets in p+p and Pb+Pb collisions at $\sqrt s = 5.02$ TeV as compared to CMS data.  Note that distributions are normalized by the number of $Z^0$ events and a kinematic cut $p_T^{\rm jet}>30 $ GeV is imposed  for the tagged jets. We observe a moderate suppression of the correlation at small $\Delta \phi_{\rm jZ}$ (large angle relative to the opposite direction of the $Z^0$ boson) in Pb+Pb relative to that in p+p collisions. This suppression is mainly caused by suppression of sub-leading jets when energy loss shifts their final transverse momentum below the $p_T^{\rm jet}=30$ GeV threshold.

To illustrate this mechanism for suppression of small angle $Z^0$+jet correlation, we also plot in Fig.~\ref{zjet_deltaphi} contributions from $Z^0$ plus only one jet (denoted as ``$Z^0+1$jet") and $Z^0$ production associated with more than 1 jets (denoted as ``$Z^0+(\ge 2)$jets")  in p+p and 0-30$\%$ central Pb+Pb collisions. We see that for $Z^0+1$jet processes, there is no significant difference between the azimuthal distributions in p+p and Pb+Pb collisions. The $Z^0$-jet correlation from for $Z^0+(\ge 2)$jets processes, however, is considerably  suppressed in Pb+Pb collisions as compared to p+p.
In $Z^0+1$jet events, the transverse momentum of $Z^0$ boson is mostly balanced by a back-to-back jet and the $Z^0$-jet azimuthal correlation is more focused  in $\Delta \phi_{\rm jZ}\sim \pi$ region where the tagged jet has a relatively large energy and is mostly a quark jet. The decorrelation of $Z^0$-jet in azimuthal angle from $Z^0+1$jet processes in this region is dominated by soft/collinear radiation, the resummation of which can be described by a Sudakov form factor. The transverse momentum broadening of this leading jet due to jet-medium interaction is negligible to that caused by soft/collinear radiation as pointed in Refs.~\cite{Chen:2016cof,Chen:2016vem,Chen:2018fqu}. This is why the contribution from  $Z^0+1$jet events to the azimuthal correlation in Pb+Pb remains almost the same as in p+p. On the other hand, the transverse momentum of $Z^0$ boson is balanced by multi-jets in $Z^0+(\ge 2)$jets processes.  The initial energy of the tagged jet is much smaller which can easily fall below  $p_T^{\rm jet} =30 $ GeV threshold due to jet energy loss.  As we can see in the comparison to the CMS data, future experimental data with much better statistics are needed to observe this suppression of small angle $Z^0$+jet correlation unambiguously.

 {\bf \textit{Summary}} ---  We have carried out a systematic study of $Z^0$+jet correlation in Pb+Pb collisions at the LHC by combining NLO matrix elements calculations with matched parton shower in Sharpa for initial $Z^0$+jet production and  Linear Boltzmann Transport model for jet propagation in the expanding QGP from 3+1D hydrodynamics.
Results from our model calculations achieve the best agreement so far with the experimental data on all four observables of $Z^0$+jet production in both p+p and Pb-Pb collisions at LHC:  azimuthal correlation in $\Delta\phi_{\rm jZ}$,  distribution of transverse momentum  imbalance $x_{\rm jZ}$, the $p_T^Z$ dependence of the mean value $ \langle x_{\rm jZ}\rangle$ and the average number of tagged jets per $Z^0$ boson $R_{\rm jZ}$.  We demonstrate the importance of both higher-order corrections and resummed soft/collinear radiation for a satisfactory description of the available experimental data on $Z^0$+jet correlations in p+p and Pb+Pb collisions. Energy loss of both leading and sub-leading jets have to be included consistently to understand the medium modifications of 
$Z^0$+jet correlations, in particular in azimuthal angle $\Delta\phi_{\rm jZ}$ and momentum imbalance  $x_{\rm jZ}$.

\textit{Acknowledgment} --- We thank E Wang, H. Zhang, P. Ru, W. Dai and S. Chen for helpful discussions. We than W. Chen for providing 3+1D hydro profile of the bulk medium in our LBT calculation. This work has been supported by NSFC of China with Project Nos. 11435004 and 11521064,  MOST of China under  2014CB845404,
NSF
 under grant No. ACI-1550228 and
 U.S. DOE under Contract No. DE-AC02-05CH11231.

\begin{acknowledgements}
\end{acknowledgements}

\end{document}